\documentclass[fullpage]{article} 
\usepackage{fullpage}
\usepackage{color}
\usepackage{authblk}
\usepackage{fancyhdr}
\usepackage{helvet}
\usepackage{amsmath}
\usepackage{psfrag}
\usepackage{setspace}
\usepackage[linktocpage]{hyperref}[2003/11/30]
\usepackage{lscape}
\usepackage{nicefrac}
\usepackage{mathrsfs}
\usepackage{units}
\usepackage{upgreek}
\usepackage{amssymb}
\usepackage{verbatim}
\usepackage[pdftex]{graphicx}
\usepackage{epstopdf} 
\usepackage{float}
\usepackage{dblfloatfix}
\usepackage{makecell}

\def\BibTeX{{\rm B\kern-.05em{\sc i\kern-.025em b}\kern-.08em
    T\kern-.1667em\lower.7ex\hbox{E}\kern-.125emX}}
\usepackage[inline]{enumitem}

\usepackage[raggedright]{titlesec}
\usepackage{cite}

\begin{document}

\title{A Schrödinger-like equation for the Thermodynamics \\of a particle in a box}
\author{A. Faigon}
\affil{Device Physics-Microelectronics Lab\\
Faculty of Engineering, University of Buenos Aires - CONICET}
\maketitle

\begin{abstract}
\noindent The particle in an expanding/contracting 1-dimension box is revisited in action-angle like variables with direct thermodynamic interpretation.
An angle dependent potential is proposed accurately describing the mechanical behavior while also capturing thermodynamic evolution -entropy production- within a canonical Hamiltonian framework. 
Heat transfer at constant volume is analyzed, and the derived thermal conductance matches the universal quantum of heat conductance $G_{Q}$ in the quantum limit.
Having a Hamiltonian scheme interpretable in thermodynamic terms, a Schrödinger-like wave equation is formulated whose wavefunction solutions contain the information about the entropy evolution. The results show exact agreement with 'classical' results for non abrupt changes.
Finally, comparisons with a  pure quantum mechanical treatment of the wave function in an expanding box confirm consistency in quasi-static regimes and reveal adiabaticity breakdown under far-from-equilibrium conditions.
\end{abstract}

\twocolumn

\section{Introduction}

Since the inception of Thermodynamics as the science of heat, considerable theoretical effort has been devoted to connecting this intriguing theory with the solid foundation that Mechanics provides to the edifice of Physics. Early pioneers such as Helmholtz, Hertz, and Boltzmann were among the first to tackle this challenge by designing mechanical models that established correspondences between mechanical and thermodynamic quantities \cite{Helmholtz 1884,Boltzmann 1884, Hertz 1910}. However, this line of inquiry was largely set aside following the success of Boltzmann’s introduction of statistical methods, later formalized by Gibbs, which led to the development of Statistical Thermodynamics. A cornerstone of this theory is the relationship between entropy and the "count of states," famously expressed in the Boltzmann equation $S=k\,ln W$ or through the ensemble formulation by Gibbs \cite{Gibbs 1902}.

Mechanical models were revisited later by de Broglie \cite{DeBroglie}, and more recently by researchers such as Gallavotti, Campisi, and García-Morales \cite{Gallavotti 1999,Campisi 2005,GarciaMorales 2008}, who emphasized the relationship between action and entropy in the form $S=k\,ln \oint pdq$. The connection between statistical and purely mechanical approaches can be understood through the loss of detailed mechanical information in the action integral, which—at least in one dimension—is represented by the enclosed phase-space volume or the number of enclosed microstates \cite{Gallavotti 1999,Campisi 2005}. %

In this work, we return to the simplest system: a particle in a one-dimensional box. We aim to develop a Hamiltonian formulation using action-angle variables with direct thermodynamic interpretations. Additionally, we extend this formulation to derive a Schrödinger-like equation for systems far from equilibrium.

The motion of a particle in an expanding box is a very interesting problem in elementary mechanics \cite{Cignoux}. Thermodynamic approaches to this problem can be found in \cite{Leff}, while quantum mechanical solutions were first proposed as early as 1969 by Doescher \cite{Doescher}. Since then, numerous studies have addressed this problem, both for its theoretical interest \cite{Berry,Cooney} and for its implications for atoms-cooling experiments \cite{Chen,DelCampo}.

This simple system serves as a testbed for the ideas developed in the following sections.

\section{Principles}

The equations of motion in Classical Mechanics relate changes in the state of motion of a system/particle with
its change in position. The energy-work theorem

\begin{equation}\label{eq:q'dp=Fdq} 
\dot q dp = F dq
\end{equation}

where $q$ and $p$ are generalized coordinates and momentum, and $F$ generalized force, states the basic form of this relationship.
Dividing \eqref{eq:q'dp=Fdq} by dt yields Newton law $F=\dot p$, which replaced in \eqref{eq:q'dp=Fdq} makes zero the differential expression
\begin{equation}\label{eq:dH=q'dp...} 
dH=\dot q dp - \dot p dq
\end{equation}

which in case F is conservative becomes an exact differential, sum of two exact differentials
\begin{equation}\label{eq:dH=dK+dV} 
dH = dK + dV
\end{equation}

the kinetic energy depending on the state of motion and the potential energy which depends on position. H is
called the hamiltonian function of the system and his value is the mechanical energy which is a constant of
motion in this case.
In a similar way, or through a Legendre transformation, we can build another differential form
\begin{equation}\label{eq:dL=pdq'+} 
dL=p d\dot q + \dot p dq
\end{equation}
sum of a kinetic term, which coincides with the kinetic energy dT in the classical case, and a positional term
which coincides as above with -dV in case $\dot p$, the force, derives from a potential V. In this case \eqref{eq:dL=pdq'+} is the exact
differential of a function $L(q,\dot q)$ the Lagrangian function of the system.
L and H are related by
\begin{equation}
L+H = p\dot q
\end{equation}

and from each of them the equations of motions are obtained.
\begin{equation}
\dot q=\frac{\partial H}{\partial p} \> and \> \dot p=-\frac{\partial H}{\partial q} ,
\end{equation}
the Hamiltonian eqs. of motion, and the Lagrangian one
\begin{equation}
\frac{d}{dt}\frac{\partial L}{\partial \dot q}-\frac{\partial L}{\partial q}=0 .
\end{equation}

Our purpose is to extend this formalism to thermodynamics by finding suitable thermodynamical
correspondences with the kinetic and positional terms above.

Consider a particle in a one-dimensional box of length $l$, the elementary thermodynamical system.
We can not observe the particle but through its effects on the walls.  Mechanical evolution for this system means adiabatic evolution -only reversible mechanical  work involved- or $p.l=constant$.
Let us explore the general thermodynamic evolutions by removing the mechanical restriction allowing the
product $p.l$, let us call it $f(p,l)\equiv p.l$, to vary. Its differential
\begin{equation} \label{eq:df}
df=p dl+l dp
\end{equation}
multiplied times $\dot q/l$, the frequency of bounce against the walls $\dot g$, yields
\begin{equation} \label{eq:a1}
\dot g df = p \dot q.dln l + \dot q dp \qquad . 
\end{equation}
Last equation is easily interpreted in pure thermodynamical quantities as
\begin{equation}
dQrev = PdV + dE\label{eq:dQrev = PdV +} 
\end{equation}
where Qrev is reversible heat, for the energy E is kinetic energy in this non-interacting particle; the reversible work P.dV term is force.dl in the one dimension box, and the force on the (quasi static) wall
is change in momentum over time $2p/(2l/\dot q)$. Alternatively, from the state equation for the ideal gas,
$PdV=kT dln Vol$ which reduces to $kT dln l$ for the one-dimension box, identifying twice the kinetic energy with temperature.

Regarding the left hand side we can rewrite
\begin{equation}\label{eq:dotg df}
\dot g df=\dot g f dln f
\end{equation}

and verify that
\begin{equation} \label{eq:kT gf}
\dot g f =p \dot q==kT
\end{equation} 
whereas
\begin{equation} \label{eq:13}
dln f =  dln \,l + dln\, p
\end{equation}
is, in k units, the contribution per degree of freedom to the entropy S of an ideal gas $k(1/3 dln Vol + 1/2 dln T)$.
Thus
\begin{equation} \label{eq:dS}
k.dln f = dS
\end{equation} 
and the l.h.s in \eqref{eq:a1} is
\begin{equation} \label{eq:TdS}
\dot g df=TdS=dQrev
\end{equation}

completing the equivalence between eqs. (\ref{eq:a1})  and \eqref{eq:dQrev = PdV +}. Eqs. \eqref{eq:dQrev = PdV +} together with \eqref{eq:TdS} constitute the fundamental equation of
Thermodynamics \cite{Callen} $TdS=dE+PdV$ relating entropy with the (un)balance between energy and mechanical work. From the last and \eqref{eq:dS} , the temperature is 
\begin{equation} \label{eq:T==dE/dS}
T=(\partial S/\partial E)^{-1}=\dot q.p
\end{equation}
consistent with our above definition in \eqref{eq:kT gf}.

\section{Evolutive Hamiltonian}

\begin{table*}[b]
\begin{center}
\begin{tabular}{||l|l|l|l||}

\hline
~ & \textbf{Mechanics} & \textbf{Evolutive} & \textbf{Thermodynamics}\\
\hline
 \textbf{Variables} & $p,q$ & $f\equiv p.l \quad,\quad dg\equiv dq/l $& $E,V$\\ 
 \hline

\textbf{Kinetic Energy} & $dK\equiv \dot{q}dp$ & $dK_{ev}\equiv\dot{g}df$ &$TdS$ \\ 

&&&\\
\textbf{Work (Potential if exact)} & $dW\equiv \dot{p}dq \quad(\equiv -d\phi) $& $dW_{ev}\equiv \dot{f}dg \quad(\equiv -d\phi_{ev})$&\\
\hline
\textbf{Energy-Work Theorem}&$\dot{q}dp=\dot{p}dq$& $\dot{g}df=\dot{f}dg$& \\
~ & & $\dot{g}df=\dot{q}dp+\dfrac{p\dot{q}}{l} dl$ & $TdS \equiv dE + PdV$ \\

~ &  ~ & \makecell[r]{($p\dot{q}\equiv kT , ln f\equiv S/k , \dfrac{p\dot{q}}{l} \equiv P  $)}& \\

&&&\\
\textbf{Hamiltonian} & $ dH(q,p)=dK+d\phi $ & $ dH_{ev}(g,f)=dK_{ev}(f)+d\phi_{ev}(g) $ & $dQ_r(S)-dQ_r(E,V)$\\
&&&\\

\hline
\textbf{Motion/Process} & inertial & adiabatic-reversible & \\
&$\dot{p}=0  $& $\dot{f}=0 $&$dS=0$\\
 & & evolutive & \\
 & &$\dot{f}\neq0 $&$\dot{S}\neq0 $\\

\hline
\end{tabular}
\caption{Mechanics, Evolutive and Thermodynamic descriptions}
\label{table:Table 1}
\end{center}
\end{table*}

Eq. \eqref{eq:TdS} is a good starting point for the identification of what will be called evolutive variables. The left hand
side is a product of the ev-velocity $\dot g$ times the differential of the ev-momentum $f$, corresponding thus to the
kinetic energy term in the mechanical hamiltonian. To complete an equivalent to the mechanical
hamiltonian scheme, an ev-positional term is required. Observing eq. \eqref{eq:dH=q'dp...}- , this term is $\dot f.dg$ the differential ev-work, equal to the ev-force $F_{ev}=\dot f$ times the differential advance or phase coordinate $g$
\begin{equation}
dg\equiv\dot gdt
\end{equation}

We can thus construct the differential
\begin{equation} \label{eq:dHev}
dHev = \dot gdf - \dot fdg
\end{equation}

the difference of the ev-energy and the ev-work, each of which becomes an exact differential if
we know the "dispersion relationship" $\dot g=\dot g(f)$, and the advance/phase dependence of the ev-force $\dot f=\dot f(g)$. In
this case eq. \ref{eq:dHev} may be written
\begin{equation}
dHev = dKev + dVev
\end{equation}

being, in light of \eqref{eq:TdS}
\begin{equation}
dKev = dQrev(f) \quad and \quad dVev = - dQrev(g) ,
\end{equation}
completing the construction of the ev-Hamiltonian, and its corresponding evolution equations
\begin{equation}
\dot g=\frac{\partial Hev}{\partial f} \quad and \quad \dot f=-\frac{\partial Hev}{\partial g} .
\end{equation}
The ev-Hamiltonian and its interpretation in Thermodynamic quantities is summarized in Table \ref{table:Table 1}.

It is worth noting at this point: If $g$ does not appear in the Hamiltonian, then $f$ is constant and can be identified with the classical Action, g is cyclic and the system does not evolve thermodynamically: dS=0. If g does appear through an ev-potential then we converted a non-conservative mechanical Hamiltonian into a conservative evolutive Hamiltonian dHev=0 from which $\dot S$ can be obtained.

In the same way, we build the ev-Lagrangian $Lev=Kev-Vev$ whose differential  is
\begin{equation}
dLev = f d\dot g + \dot fdg \quad,
\end{equation}
from which we obtain the Lagragian equation of evolution
\begin{equation}
d/dt(\partial Lev/\partial \dot g) - \partial Lev/\partial g =0 \quad.
\end{equation}

Finally, for completeness,we have
\begin{equation}\label{eq:Lev+Hev}
Lev+Hev = f \dot g=kT \quad.
\end{equation}

\section{Change of volume}

Let us build the ev-Hamiltonian for a particle moving in a one-dimensional box that expands or contracts at a rate $\dot l$. From \eqref{eq:kT gf}, the kinetic term is
\begin{equation}
dKev = dQrev(f) = \dot g df.
\end{equation}

The potential term , accordingly, is the r.h.s. of \eqref{eq:dQrev = PdV +} in terms of g
\begin{equation}
dVev = -dQrev(g) = -(dE+PdV)/dg . dg.
\end{equation}

For the system under consideration, we have
\begin{equation}
dE =-dL=-F.dl=-\dfrac{\Delta p}{\Delta t}.\dfrac{\dot l}{2}dt=-(2p-m\dot l) \dfrac{\dot l}{2}dg
\end{equation}

using $dt=dg/\dot g$, for $\Delta p$ is the momentum change on each bounce and $\dot g$ the frequency of bounce occurrence.
And the quasi-static work PdV is
\begin{equation}
F_{st} . dl = 2p \dfrac{\dot l}{2}dg
\end{equation}

where $ F_{st}$ is the force on a static wall. Thus
\begin{equation}
dVev = -\dfrac{1}{2} m \dot l^{2} dg ,
\end{equation}

and
\begin{equation}\label{eq:dHev2}
dHev(f,g)= \dfrac{f}{ml^{2}} df -\dfrac{1}{2} m \dot l^{2} dg .
\end{equation}

From this we have

\begin{equation}\label{eq:f'deHev}
 \dot f=-\frac{\partial Hev}{\partial g} = \dfrac{1}{2} m \dot l^{2} ,
\end{equation}
which is always positive, and so is the rate of entropy production
\begin{equation}
\dot S = k \dot f/f >= 0
\end{equation}

whether in compression or expansion processes.
The expression for $\dot f$  in \eqref{eq:f'deHev} obtained from the Hamiltonian framework agrees with that derived directly from the definition of f: Using $p_g=m\dot l (\alpha-g)$ for the momentum after the $g$ bounce, $\alpha \equiv p_o/m\dot l$ and $l[g] =lo \dfrac{\alpha}{\alpha-g}$ -see Appendix A-, we get

\begin{equation}\label{eq:f'_g} 
\dot f=\dfrac{\Delta f}{\Delta g}\dot g=
\dfrac{f_{g+1/2}-f_{g-1/2}}{1} \dfrac{p_g/m}{\tiny{\dfrac{1}{2}}(l_{g+1/2}+l_{g-1/2})}=\dfrac{1}{2}m\dot {l}^2
\end{equation}

validating the ev-hamiltonian formulation.

\section{Heat transmission at constant volume}

Let us show the Lagrangian
\begin{equation}\label{eq:Levconstvol} 
Lev(g,\dot g) = \dfrac{1}{2}ml^{2}\dot g^{2}-\dfrac{k}{2}(Te-To)e^{-c(g-go)}\quad,
\end{equation}
$c$ being a constant, describes the evolution at constant volume for the system initially at temperature $To$ immersed at $g=go$ in a thermal bath with temperature $Te$. The Euler-Lagrange equation for $g$ becomes
\begin{equation}\label{eq:ml^2.g''-}
ml^{2} \ddot g-\dfrac{k}{2}c(Te-To)e^{-c(g-go)}=0\quad.
\end{equation}
Solving for $\dot g^2$ and using the identity $f\dot g=kT$ together with the easy to verify velocity-momentum relationship $f(\dot{g})=\dot{g}/(m l^2)$ we get the Thermodynamic interpretation 
\begin{equation}\label{eq:T-To=}
T-To=(Te-To)[1-e^{-c(g-go)}]\quad.
\end{equation}

The conductance involved in the process is  $\sigma \equiv 1/(Te-T).dQ/dt$, where $dQ$ is the heat exchanged through one wall in time $dt$ corresponding to $dg$, with $dt = dg/\dot g$. Using dQ=Cv.dT and T from \eqref{eq:T-To=}, the conductance is
\begin{equation}\label{eq:sigma=c...}
\sigma=c Cv  \dot g=c\dfrac{Cv  kT}{f}\leq c \dfrac{ k^{2}T}{h}
\end{equation}
The expresion on the r.h.s  is the maximum value for the conductance, corresponding to $Cv=k/2$ for one degree of freedom, and the minimum value for $f$, $f=h/2$. It has the same form  of the universal quantum of heat conductance predicted by Rego and measured by Roukes et al. [12,13]. Identical expression is obtained by giving  the factor $c$ the value $\pi^{2}/3$.

\section{Non equilibrium and ev-Schrödinger Equation}

The above description assumes slow changes, i.e. small relative changes along one g or period: $\delta f/f<<1$. It is this condition that enables a trajectory representation $g(t)$ as assumed up to this point. If it fails, a wave-probabilistic approach is suggested in line with the transition from classical to quantum mechanics. I.e. we are requiring a Schrodinger-like wave equation

\begin{equation}\label{eq:Hev^Psi=Eev.Psi}
\hat{H}ev \Psi=Eev \Psi
\end{equation}

For the former case of change of volume, from eq. \eqref{eq:dHev2}, integrating from $(g=0, f=fo)$ to $(g,f)$,

\begin{equation}\label{eq:Hev=}
Hev=\dfrac{f^{2}-fo^2}{2ml^{2}}+ (\alpha-g) \dfrac{1}{2} m \dot l^{2} 
\end{equation}

where we set the initial value for the ev-potential to $Vev[g=0] = \alpha \frac{1}{2} m \dot l^{2}$ which is  the amount of energy available to be converted in ev-kinetic energy $TdS$ as this conversion occurs at a rate of $\frac{1}{2} m \dot l^{2}$ each $g$ and the evolution in expansion exhausts  ($v<\dot{l}/2)$ after $\alpha$ bounces ($g=\alpha$).

\begin{figure}[t] 
\begin{center}
\includegraphics[width=7 cm]{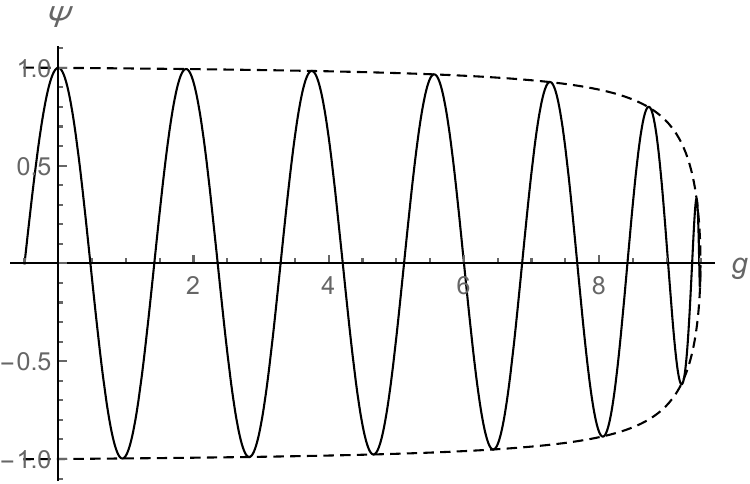}
\caption{$\Psi(g)$ for $\alpha=10$. Envelopes, in dashed lines, are normalized $\:^{+}_{-} f^{-1/2}$
}\label{fig:psi}
\end{center}
\end{figure}

\begin{figure}[t]
\begin{center}
\includegraphics[width=9
 cm]{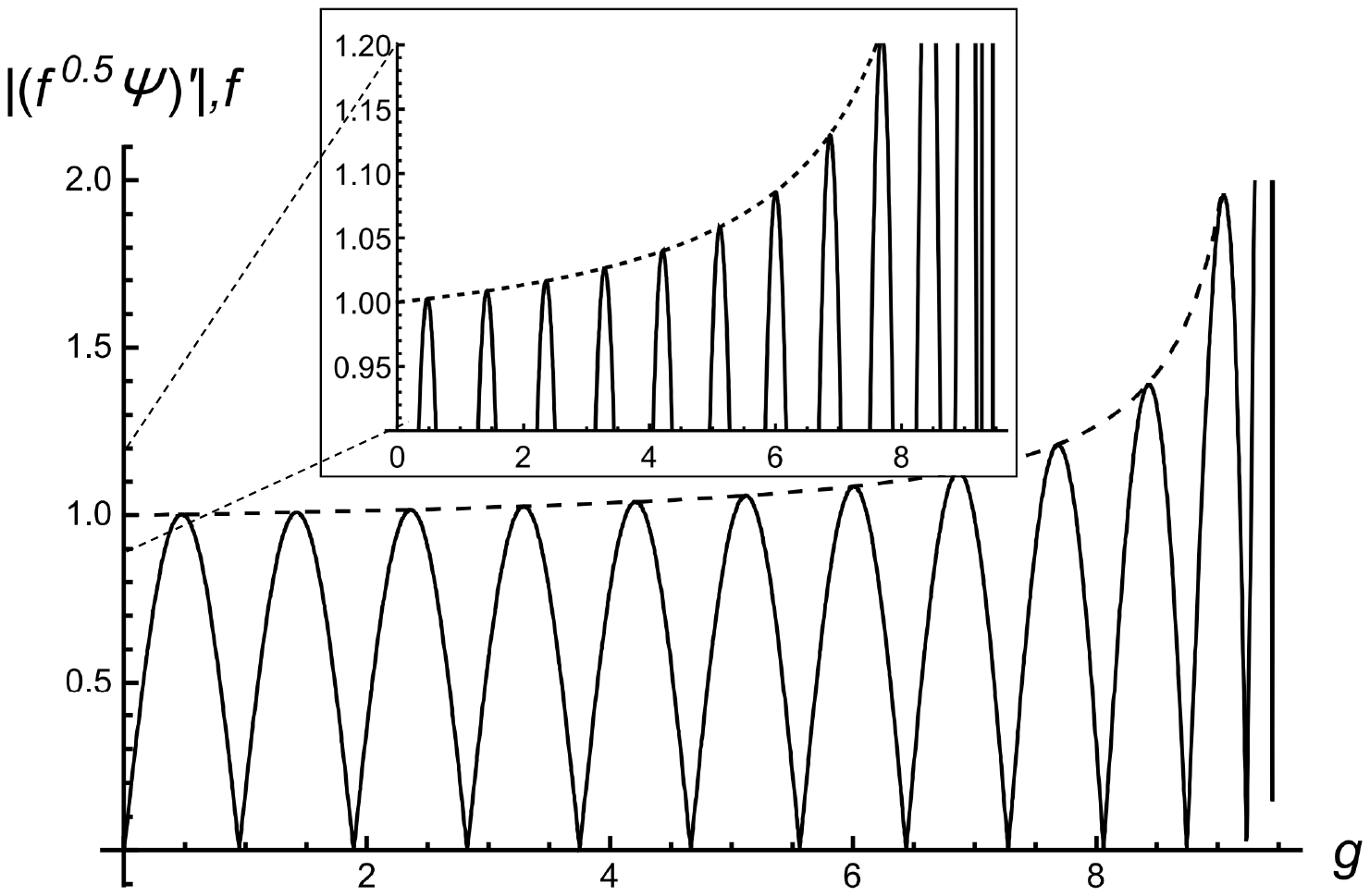}
\caption{Derivative of $f^{0.5}\Psi$ and f(g)(dashed line)}
\label{fig:psi'& f}
\end{center}
\end{figure}

Using the momentum operator 
\begin{equation}\label{eq:f^}
\hat{f}=-i\hslash \dfrac{\partial}{\partial g},
\end{equation}

the following wave equation is obtained 

\begin{equation}\label{eq:Waveeq} 
\frac{\partial^{2} \Psi(g)}{\partial g^{2}}+\dfrac{fo^2}{\hslash^{2}}(\dfrac{\alpha-g+1/2}{\alpha-g}-1) \Psi(g)=0\:
\end{equation}

The analytical solution is a polynomial expression involving Whittaker functions. Fig. \ref{fig:psi} presents a graphical representation of the numerical solution. The wave function collapses to zero at $g=\alpha$ as after the last bounce the particle classical velocity is lower than that of the wall. It ceases all interaction with the outside world, it is no longer observable. 
 The wave representation was validated by comparing the wave number k(g) (multiplied by $\hbar$) with the classical momentum f(g), which should coincide at  sufficiently slow rates of volume change. In Fig. \ref{fig:psi} we plotted the amplitude of the wave function, which, following the WKB representation, scales as $k^{-0.5}$,  and is well fitted by the classical expression $f^{-0.5}$. A more detailed comparison is made by extracting the wave number from the exponent of the WKB representation. This is achieved using the derivative of the exact solution multiplied by $f^{0.5}$, $d/dg (f^{0.5}\Psi)$, as shown in Figure \ref{fig:psi'& f}, alongside the classical value of the momentum f. The inset provides a close-up view of this comparison, highlighting the accuracy of the correspondence between the two magnitudes.

\begin{figure}[H]
\begin{center}
\includegraphics[width=7 cm]{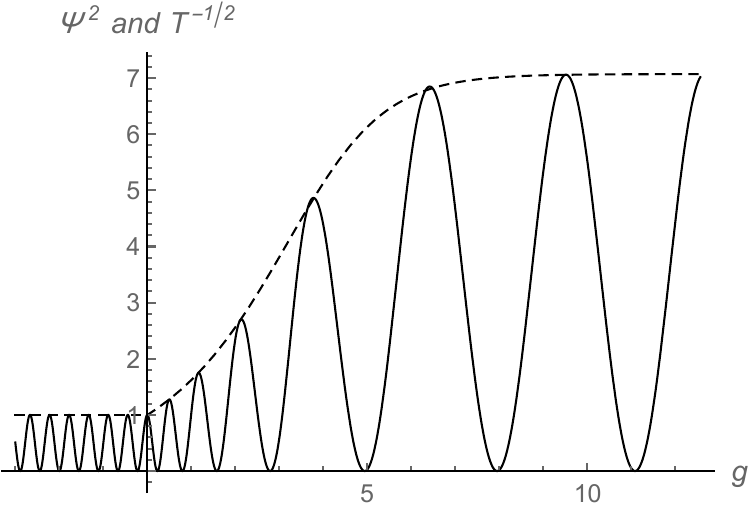}
\caption{Probability density and $T^{-0.5}$ (dashed) for the evolution from $50 Te$ to $Te$ in normalized units.}
\label{fig:fig psi vol cte}
\end{center}
\end{figure}

For the constant volume evolution, the Hamiltonian is, from \eqref{eq:Levconstvol} and \eqref{eq:Lev+Hev},
\begin{equation}\label{eq:HevVolCte}
Hev=\dfrac{f^{2}}{2ml^{2}}+\dfrac{k}{2} (Te-To) e ^{-c (g-go)} 
\end{equation}
and the corresponding wave-equation is
\begin{equation}\label{eq:WaveeqVolCte} 
-\dfrac{\hslash^{2}}{2 m l^{2}} \frac{\partial^{2} \Psi(g)}{\partial g^{2}}+\dfrac{k}{2} (Te-To) e ^{-c (g-go)} \Psi(g)=Eev \Psi(g)\qquad.
\end{equation}
Fig \ref{fig:fig psi vol cte} illustrates the evolution at constant volume from $50 Te$ to $Te$.

Again the Figure shows correspondence with the "classical" treatment this time through the amplitude of $\psi^2$ being proportional to $f^{-1}$ or $T^{-1/2}$. 

\section{Back to Mechanics}
\begin{figure}[H]
\begin{center}
\includegraphics[width=8 cm]{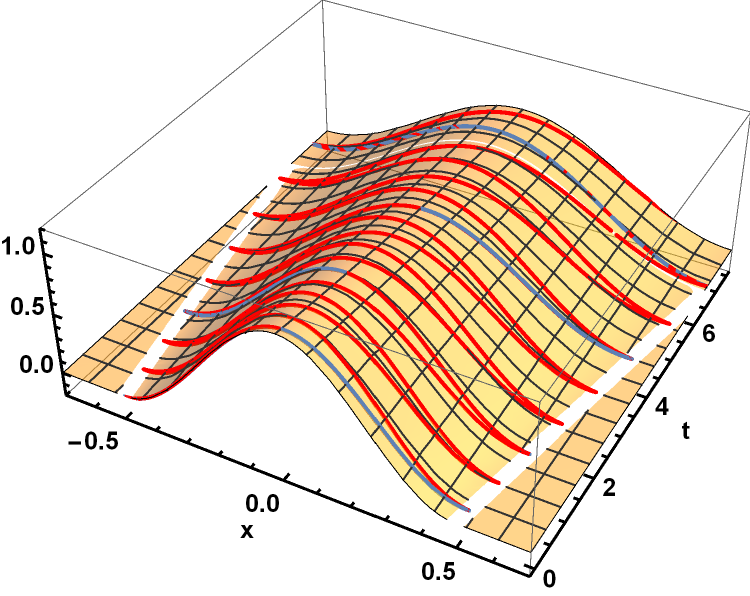}
\caption{The surface is the solution $\Psi_{D}(x,t)$ found in \cite{Doescher}, in red the same solution along the particle classical path $\Psi_{D}(x(g),t(g))$, in blue our solution $\Psi(x(g),t(g))$.}
\label{fig:fig_Doesch_alf=50}
\end{center}
\end{figure}

If the above formalism is correct, we should be able to go back and reconstruct from it the $\Psi(x,t)$ wave function evolution as results from straight quantum mechanics treatment as first done in \cite{Doescher}. The way to compare our $\Psi(g)$ with the $\Psi(x,t)$ is through the dependence x(t) along a posible particle classical path which between bounces is $x(t) = 1/2 .l_{g-1/2}+ v_{g}.t$. Thus we can compare 
 $\Psi(x(g),t(g))$ 
 with the solution found by Doescher \cite{Doescher} $\Psi_{D}(x,t)]$ . Fig \ref{fig:fig_Doesch_alf=50} shows the result of this comparison for a very slow expansion, $\alpha$=50, from  $l_{0}$=1 to  $l_{f}$= 1.4.
 
\begin{figure}[H]
\begin{center}
\includegraphics[width=8 cm]{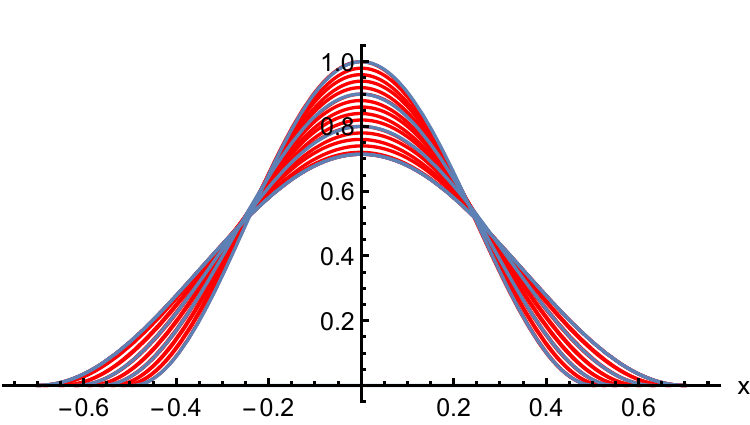}
\caption{Projections of the curves of Fig \ref{fig:fig_Doesch_alf=50}  on the plane t=0. Only some of the blue curves were plotted for clarity.}
\label{fig:/fig_Doesch_alf=50_projection}
\end{center}
\end{figure}

Projections of the solutions on the plane t=0 show exact agreement -Fig. \ref{fig:/fig_Doesch_alf=50_projection}-

\begin{figure}[H]
\begin{center}
\includegraphics[width=8 cm]{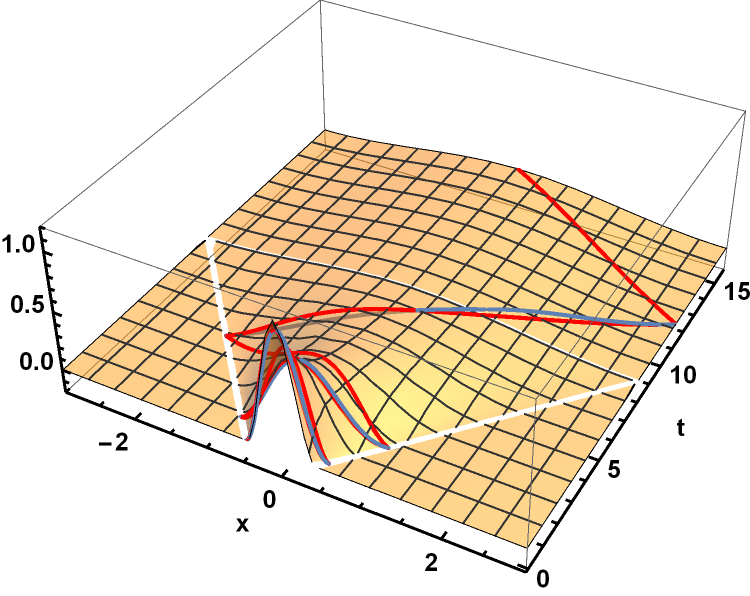}
\caption{Same as Fig \ref{fig:fig_Doesch_alf=50} for $\alpha$=5 and $l_{f}$=6.}
\label{fig:fig_Doesch_alf=5}
\end{center}
\end{figure}

As the expansion rate increases, red and blue curves start to diverge because Doescher's treatment assumes an adiabatic process, $E$ proportional to $1/l^2$ , eq. 4 in \cite{Doescher}, or $f$ constant. An example is shown in Fig. \ref{fig:fig_Doesch_alf=5}

 \begin{figure}[H]
\begin{center}
\includegraphics[width=8 cm]{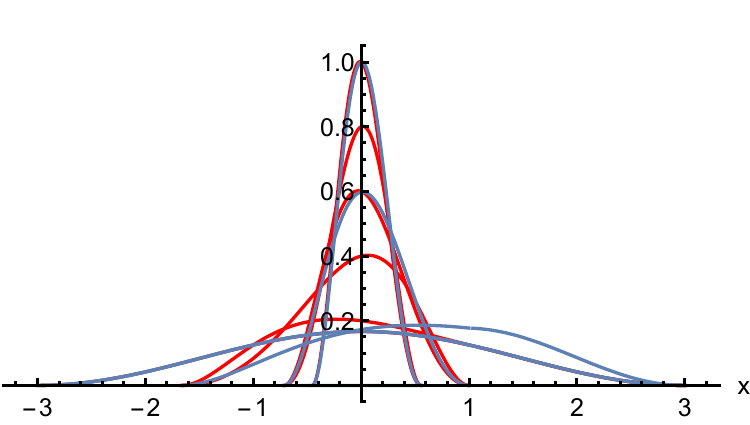}
\caption{Projections to t=0 for $\alpha$=5 and $l_{f}$=6.}
\label{fig:fig_Doesch_alf=5_projection}
\end{center}
\end{figure}

\begin{figure}[H]
\begin{center}
\includegraphics[width=8 cm]{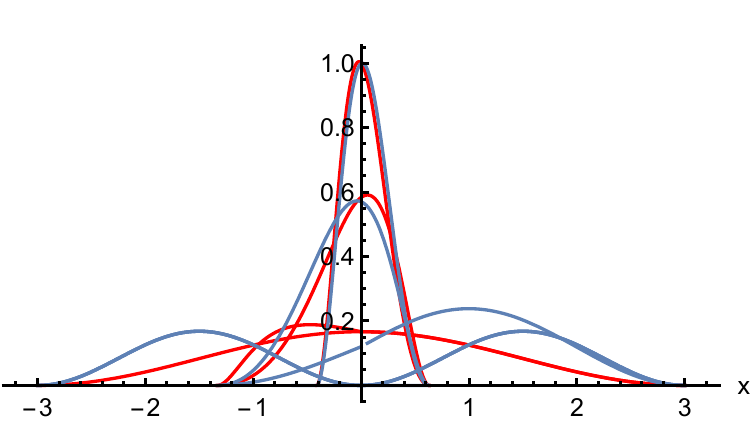}
\caption{Projection to t=0 for $\alpha$=2.4 and $l_{f}$=6. Evolution in red shows the particle remains in the same lowest state forced by the model, blue curve for the final length $l_{f}$=6 shows the particle in the next quantum state}
\label{fig:fig Doesch alf=2.4 projection}
\end{center}
\end{figure}

Finally, the breakdown of adiabaticity becomes evident as the expansion proceeds at an even faster rate as illustrated in Fig. \ref{fig:fig Doesch alf=2.4 projection},showing a transition to a different quantum state.

\section{Helmholtz}
Helmhotz proposed the first canonical scheme for thermodynamics. He assumed the existence of a quantity $\epsilon$ whose time derivative is the temperature. Thus, if $\varepsilon$ is the generalized force associated to $\epsilon$ then
\begin{equation}\label{eq:helmholtz1} 
dE=\varepsilon d\epsilon-p dV=\varepsilon \dot \epsilon dt - pdV,
\end{equation}
or
\[\varepsilon \dot \epsilon dt=TdS,\]
or
\[\dot S=\varepsilon\]
allowing a lagrangian formulation with $S$ the generalized momentum conjugate to $\epsilon$.\cite{DeBroglie2}

From the above development, it becomes clear that Helmholtz's elusive quantity is

\begin{equation}
d\epsilon = f. dg/k=p.dq/k
\end{equation}

the reduced or Maupertius action whose conjugate momentum is 

\begin{equation}
ds =dS= k dln f.
\end{equation}

By their definitions the new coordinates action-entropy preserve the elemental phase space volume

\begin{equation}
ds.d\epsilon = df.dg
\end{equation}

ensuring the canonicalicity of the transformation.
The "work-energy" theorem translates in

\begin{equation}
TdS-\dot{s}d\epsilon=0
\end{equation}

which may be the basis of a new hamiltonian scheme if the second term of last equation can be written as $\partial V(\epsilon)/\partial \epsilon .d\epsilon$

\section{Summary}

This work revisits the elementary system consisting of a particle in a one-dimensional box undergoing expansion or contraction, introducing a Hamiltonian framework with direct thermodynamic interpretation. 
In these non-adiabatic processes, the adiabatic invariant f=p.l does vary. It plays the role of momentum in the hamiltonian and its variation is directly related with the entropy production as is the kinetic part of the hamitonian identified with the reversible heat. This lead us to call the whole scheme as an evolutive Hamiltonian.

Heat exchange at constant volume was also treated within the same formalism. The proposed Lagrangian leads to a calculated thermal conductance matching the universal quantum of heat conductance , illustrating the applicability of the framework to quantum thermal problems.

The established correspondence between mechanical and evolutive coordinates within the Hamiltonian scheme allowed the derivation of a Schrödinger-like wave equation capable of capturing the system's behavior beyond quasi-static conditions. Solutions demonstrate correspondence with classical results in the WKB approximation and reveal adiabaticity breakdown in far-from-equilibrium scenarios.
Comparisons with previous quantum mechanical solutions, such as Doescher's, confirm consistency in quasi-static regimes while highlighting deviations at higher expansion rates.

This work could contribute to the understanding of the interplay between mechanical and thermodynamic systems and provides a new -to the best of our knowledge- framework for exploring far-from-equilibrium phenomena. 

Extension of these principles to higher dimensional systems is straightforward for non interacting degrees of freedom. Future research could extend them to more complex interactions.

\appendix
\section{Appendix A}
At t=0 the box has a length $lo$ and expands symetrically -to avoid considering the motion of the center of mass- at a rate $\dot l$; the particle at the center of the box moves to the right with velocity $vo=\alpha \dot l$. The advance parameter $g$ counts the number of box lengths traveled, so the particle reachs the right wall at $g=1/2$ in time
\begin{equation}
\delta t_{1/2}=\dfrac{lo/2}{(vo-\dot l/2)},
\end{equation}
the box length having changed by
\begin{equation}
\dfrac{l_{1/2}}{lo} = 1 + \dfrac{1}{2} \dfrac{1}{\alpha-1/2}.
\end{equation}

After the bounce the particle returns to the center of the box in a time
\begin{equation}
\delta t_{1}=\dfrac {(lo+\dot l \delta t_{1/2})/2}{(v_{1}-\dot l/2)},
\end{equation}

where $v_{g}$ is the particle velocity after g bounces
\begin{equation}
v_{g}=vo-g.\dot l,
\end{equation}

being the box length incremented by
\begin{equation}
\dfrac{l_{1}}{l_{1/2}} = 1 + \dfrac{1}{2} \dfrac{1}{\alpha-1}.
\end{equation}

The box length when the particle passes exactly through its center after g bounces is, recursively,
\begin{equation}
\dfrac{l_{g}}{lo}=\dfrac{l_{g}}{l_{g-1/2}}.\dfrac{l_{g-1/2}}{l_{g-1}}....=\prod_{n=1}^{2g} (1+\dfrac{1}{2} \dfrac{1}{\alpha-n/2})=\dfrac{\alpha}{\alpha-g}
\end{equation}
The r.h.s. holds if the integer and half integer subindexes -corresponding to particle at the center and particle at the wall
respectively- are replaced by a continuos dependence on $g$ in the l.h.s. yielding the continuous function
\begin{equation}
l[g] =lo \dfrac{\alpha}{\alpha-g} .
\end{equation}

\bibliographystyle{phaip}

\end{document}